\documentclass{aa}  
\usepackage{graphicx}
\usepackage{txfonts}
\usepackage{color}
\begin{document} 

   \title{Dynamical friction in the quasi-linear formulation of  modified Newtonian dynamics (QuMOND)}
\titlerunning{Dynamical friction in QuMOND}
   \author{Pierfrancesco Di Cintio \inst{1,2,3}
           \and
          Federico Re \inst{4,5}
           \and
         Caterina Chiari \inst{6,7}
          }
   \institute{CNR-ISC, via Madonna del Piano 17 50022 Sesto Fiorentino, Italy
   \and
              INAF-Osservatorio Astronomico di Arcetri, Largo Enrico Fermi 5 50125 Firenze Italy
   \and          
              Dipartimento di Fisica e Astronomia, Universit\'a di Firenze \& INFN-Sezione di Firenze, via Sansone 1 50022 Sesto Fiorentino, Italy\\
              \email{pierfrancesco.dicintio@unifi.it}
   \and          
              Dipartimento di Fisica ''Giuseppe Occhialini", Universit\'a di Milano Bicocca, Piazza della Scienza 3 20126, Milano, Italy 
   \and
              INFN-Sezione di Milano Via Celoria 15 20133, Milano, Italy\\
              \email{federico.re@unimib.it}            
   \and
              Dipartimento di Scienze Fisiche, Informatiche e Matematiche, Universit\'a di Modena e Reggio Emilia, Via Campi 213/A, I-41125 Modena, Italy
              \and
             CNR-NANO, via Campi 213/A I-41125, Modena, Italy\\
             \email{caterina.chiari@unimore.it}
             }
   \date{Received ??; accepted ??}
%%%%%%%%%%%%%%%%%%%%%%%%%% 
  \abstract
   {}
   {We explore the dynamical friction on a test mass in gravitational systems in the quasi-linear formulation of modified Newtonian dynamics (QuMOND).}
   {Exploiting the quasi-linearity of QuMOND, we derived a simple expression for the dynamical friction in akin to its Newtonian counterpart in the standard Chandrasekhar derivation. Moreover, adopting a mean field approach based on the Liouville equation, we were able to obtain a more rigorous (albeit in integral form) dynamical friction formula that can be evaluated numerically for a given choice of the QuMOND interpolation function.}
   {We find that our results are consistent with those of previous works. We observe that the dynamical friction is stronger in MOND with respect to a baryon-only Newtonian system with the same mass distribution. This amounts to a correction of the Coulomb logarithmic factor via additional terms that are proportional to the MOND radius of the system. Moreover, with the aid of simple numerical experiments, we confirm our theoretical predictions and those of previous works based on MOND.}
   {}
%%%%%%%%%%%%%%%%%%%%%%%%%%%%%%%%%%
   \maketitle
%
%-------------------------------------------------------------------
\section{Introduction}
Modified Newtonian dynamics (hereafter MOND, \citealt{1983ApJ...270..365M}) is an alternative theory of (classical) gravity that was originally introduced to solve the missing mass problem on astrophysical and cosmological scales without resorting to dark matter theory (hereafter DM). In a Lagrangian formulation (see \citealt{1984ApJ...286....7B}), MOND can be amended to a substitution of the Poisson equation relating the gravitational potential, $\Phi,$ and mass density, $\rho$, with the non-linear field equation:
\begin{equation}
\label{MOND}
    \vec{\nabla}\cdot\left[\mu\left(\frac{||\vec{\nabla}\Phi||}{a_0} \right)\vec{\nabla}\Phi\right]=4\pi G\rho,
\end{equation}%
where the acceleration, scaling as $a_0\approx 1.2\times 10^{-2}{\rm ms}^{-2}$, is a new universal constant; whereas the  interpolating function, $\mu,$ (which is, in principle, unknown) is monotonic with the asymptotic behaviour, as follows:\ 
\begin{equation}
\label{limits}
    \mu(x)\rightarrow^{x\gg1}1, \quad \mu(x)\sim^{x\ll1}x.
\end{equation}
A widely adopted form used for $\mu$ is:  
\begin{equation}\label{interpmu}
    \mu(x)=\frac{x}{1+x}.
\end{equation}
According to the equations above, for $\nabla\Phi\gg a_0$, we can  recover the Newtonian limit and Eq. (\ref{MOND}) reduces to the Poisson equation. Vice versa, for $\nabla\Phi\ll a_0$ the system is in the so-called deep-MOND limit (hereafter, dMOND) and Eq. (\ref{MOND}) becomes the p-Laplace equation:
\begin{equation}
\label{dMOND}
\vec{\nabla}\cdot\left({||\vec{\nabla}\Phi||} \vec{\nabla}\Phi\right)=4\pi a_0G\rho.
\end{equation}
For a given mass density, $\rho,$ the right-hand-side of the Poisson equation and Eq. (\ref{MOND}) are the same. We can therefore eliminate $\rho$ , thereby obtaining the relation:
\begin{equation}
\label{gmgn}
  \mu\left(\frac{||\mathbf{g}_M||}{a_0}\right)\mathbf{g}_M=\mathbf{g}_N+\mathbf{S},
\end{equation}%
between the MOND and Newtonian force fields $\mathbf{g}_M$ and $\mathbf{g}_N$, where $\mathbf{S}\equiv\nabla\times\mathbf{h}(\rho)$ is a density-dependent solenoidal field that zeros-out for systems in spherical, cylindrical, or planar symmetry, while it is typically non-zero for more general density profiles.\\
\indent MOND has been rather successful in reproducing the kinematics of galaxies without the need for DM to be included (\citealt{Bugg_2015,2023MNRAS.519.4479Z}). In fact, it has proven to be a valid alternative to the $\Lambda$CDM paradigm where the latter is challenged, across a broad range of gravitational problems, such as the structural properties of dwarf galaxies (\citealt{1994ApJ...429..540M,2007ApJ...667..878S,2013ApJ...766...22M,2022MNRAS.515.2981A});  radial acceleration relation of galaxies (RAR, \citealt{2017ApJ...836..152L,Mistele_2024,2024arXiv240609685M});  profiles of low-surface-brightness galaxies (\citealt{Sanders_2021});  dynamics of wide binary stars (\citealt{2012EPJC...72.1884H,10.1093/mnras/stad3446,2024ApJ...960..114C});  evolution of globular clusters of ultra-diffuse galaxies (\citealt{B_lek_2021}); and the orbital velocity of interacting galaxy pairs (\citealt{2022MNRAS.512..544S}).\\
\indent MOND has been extensively investigated numerically (\citealt{1999ApJ...519..590B,2004MNRAS.347.1055K,2007A&A...464..517T,2007ApJ...660..256N,2007MNRAS.381L.104N,2007MNRAS.379..597N,2008MNRAS.386.1588S,2009MNRAS.396..109W,2009ApJ...694.1220M,2021MNRAS.503.2833R,2022MNRAS.517.3613K, Banik_2022, 2023MNRAS.519.5128N,2023A&A...678A.110R,2023MNRAS.524.5291S}), for collisionless mean field processes; for example: violent relaxation and phase-mixing in monolithic collapse, galaxy merging, and the vertical dynamics of disk galaxies. Unfortunately, due to the non-linearity of the theory (cfr. Eq. \ref{MOND}) and, hence, the absence of the superposition principle, there is much less information available on the collisional processes (see e.g. \citealt{B_lek_2021}). For example, \cite{2004MNRAS.351..285C} used an approach based on the fluctuations of uniform field to estimate the MOND two-body relaxation ($t_{2b}$, \citealt{1941ApJ....93..285C,1941ApJ....93..323C}) and dynamical friction ($t_{DF}$, \citealt{1943ApJ....97..255C,1943ApJ....97..263C,1943ApJ....98...54C}) time scales in the dMOND limit. The outcome of this study (see also the numerical simulations in \citealt{2008MNRAS.386.2194N}) shows that a test mass, $M$, crossing a purely baryonic system would feel a stronger dynamical friction (hereafter, DF) force due to the encounters with the system's stars in MOND than in the parent Newtonian system without DM; however, it would undergo an only slightly more efficient DF than in the equivalent Newtonian system (ENS). The is comprised of the baryonic plus DM system constructed such that the potential is the same as the purely baryonic MOND model. The Newtonian and MOND DF time scales $t^{\rm N}_{\rm DF}$ and $t^{\rm M}_{\rm DF}$, are related as:
\begin{equation}\label{tmond}
t^{\rm M}_{\rm DF}=\frac{\sqrt{2}}{1+\mathcal{R}}t^{\rm N}_{\rm DF},
\end{equation}
where $\mathcal{R}$ is the ratio of the amount of DM to baryons in the ENS.\\
\indent In MOND, the collisional direct $N-$body simulations are intrinsically unfeasible; thus, results such as those given by Eq. (\ref{tmond}) cannot be exhaustively explored via numerical experiments. The reason for this is the absence of a MONDian expression for the force exerted by a point-like particle. To address this issue, \cite{1986ApJ...302..617M} proposed the approximated expression for the force exchanged between two masses $m_1$ and $m_2$ placed at distance, $r$,
\begin{equation}
\label{f12}
F_{1,2}\approx\frac{m_1m_2}{\sqrt{m_1+m_2}}\frac{\sqrt{Ga_0}}{r}.
\end{equation}
The latter is however valid only in the far field limit, thus making it unusable in a direct $N-$body code, as it lacks a regime bridging to the Newtonian $1/r^2$ limit.\\  
\indent More recently, \cite{2010MNRAS.403..886M} formulated a quasi-linear MOND theory (hereafter QuMOND), where the governing field equation is:
\begin{equation}
\label{QuMOND}   \vec{\nabla}\cdot\left[\nu\left(\frac{||\vec{\nabla}\Phi^N||}{a_0} \right)\vec{\nabla}\Phi^N\right]=\Delta\Phi^M,
\end{equation}%
formally identical to Eq. (\ref{MOND}) where now, $\Phi^N$ is the Newtonian potential generated by the density $\rho$ through the usual Poisson equation:
\begin{equation}
\label{Newton}
    \Delta\Phi^N=-\vec{\nabla}\cdot\vec{g}^N=4\pi G\rho,
\end{equation}%
and the new interpolating function $\nu(y)$ is related to $\mu(x)$ via:
\begin{equation}
\label{mu nu}
    \begin{cases}
        x=y\nu(y) \\
        y=x\mu(x)
    \end{cases} \Rightarrow \mu(x)=\frac{y}{x}=\frac{1}{\nu(y)}.
\end{equation}
The asymptotic behaviours for $\nu$ in the Newtonian and in the dMOND limits are
\begin{equation}
\label{Qu limits}
    \nu(y)\rightarrow^{y\gg1}1, \quad \nu(y)\sim^{y\ll1}y^{-1/2}.
\end{equation}
We note that Eq. (\ref{QuMOND}) can be rearranged as
\begin{equation}\label{rhoaug}
\tilde\rho=-\frac{1}{4\pi G}\nabla\cdot[\nu(g^N/a_0)\mathbf{g}^N];\quad \mathbf{g}^N=-\nabla\Phi^N,
\end{equation}
that is, the QuMOND potential satisfies the Poisson equation for the auxiliary density $\tilde\rho$. In practice, in QuMOND, we have to evaluate the Newtonian potential $\Phi^N$ generated by $\rho$, evaluate its gradient and then via a non-linear algebraic step obtain the auxiliary MOND density $\tilde{\rho}$ via Eq. (\ref{rhoaug}) that acts as source for the potential $\Phi^M$.\\
\indent At variance with the usual MOND formulation, in its quasi-linear formulation the interpolating function $\nu$ has a stronger effect on the form of the gravitational potential than its counterpart $\mu$ in  Eq. (\ref{MOND}).
For the usual choice (\ref{interpmu}), we obtain: 
\begin{align}
    \nu(y)=\frac{1}{2}+\frac{1}{2}\sqrt{1+\frac{4}{y}}. \label{usual nu}
\end{align}
We note that QuMOND contains a sort of superposition principle for the auxiliary densities $\tilde{\rho}$. It is therefore tempting to try to derive an expression for the DF force in this context, making use of purely kinetic arguments as in the original derivation. In this paper, we first discuss a simple expression of the QuMOND DF summing up the contributions of the effective densities for point-particles of mass, $m$, interacting with a test mass, with some simple choices for $\nu$. Moreover, we extend the mean field formulation of DF, originally derived in Newtonian gravity by \cite{1983Ap&SS..97..435K} to the case of QuMOND and discuss the results of simple numerical experiments.  

\section{Dynamical friction: Classical approach}

\subsection{Newtonian case}\label{newtonian}
Before tackling the QuMOND DF formula, we start by briefly recalling the ingredients  of the derivation of its Newtonian expression (for a more detailed review, see e.g. \citealt{binney} or \citealt{2021isd..book.....C}). We consider a test mass, $M,$ travelling at $v_M$ through an infinite background medium of particles of a mass, $m,$ and (by now constant) number density, $n$, with velocity distribution function, $f(v)$. In each encounter, $M$ and a background particle exchange a force expressed as $F\approx{GMm}/{b^2}$, acting for an interval of time, $\Delta t\approx{b}/{v_M}$, where $b$ is the impact parameter of the dynamical collision. The associated variation of momentum is $\Delta p\approx F\Delta t\approx{GMm}/{v_M b}$. The rate of collisions of $M$ is then
\begin{equation}
    \dot{N}_{\rm enc}=\frac{2\pi nb db dx}{dt}=2\pi n v_M b db.
\end{equation}%
From this, the velocity diffusion coefficient along the trajectory of $M$ can be written as:
\begin{equation}
\label{Newt Dv}
    D_v=\int \Delta p^2 \frac{dN_{\rm enc}}{dt}=2\pi G^2 M^2\frac{m^2 n}{v_M}\int\frac{db}{b}.
\end{equation}
In the equation above, the integral in the impact parameter $b$ diverges for both $b\rightarrow0$ and $b\rightarrow\infty$ (infrared and ultraviolet divergence, respectively). It is however possible to impose suitable cut-offs $b_{\rm min}$ and $b_{\rm max}$. This factor, after integration takes the form of the Coulomb logarithm (in analogy with the same quantity in plasma physics) $\ln\Lambda$, where $\Lambda=b_{\rm max}/b_{\rm min}$.\\
\indent Let us now assume that $f(v)$, defined such that 
\begin{equation}
\int f(\vec{x}, \vec{v})d^3\vec{x} d^3\vec{v}=N
\end{equation}
is the total number of particles, is position independent and isotropic for all velocities. For such a system, the DF force per unit mass acting on $M$ is given as
\begin{align}
    \frac{d\mathbf{v}_M}{dt}&=-16\pi^2G^2n(M+m)m\ln\Lambda\frac{\Psi(v_M)}{v_M^3}\vec{v}_M, \label{Chandra}
\end{align}
where
\begin{equation}
\Psi(v_M)=\int_0^{v_M}v^2f(v)dv
\end{equation}
is the so-called fractional velocity volume function (\citealt{2021isd..book.....C}). Therefore, in the assumption of velocity isotropy, only particles moving slower than $M$ contribute to its slowing down. Moreover for $M\gg m$, in Eq. (\ref{Chandra}) the factor $m(M+m)n$ becomes $M\rho$, where $\rho=nm$ is the background mass density. In Newtonian systems with DM, two species of particles exert DF on the test mass $M$. Following \cite{2010AIPC.1242..117C}, Eq. (\ref{Chandra}) can be rewritten for a model with a mass spectrum as 
\begin{align}
    \frac{d\mathbf{v}_M}{dt}&=-16\pi^2G^2n(M+\langle m\rangle)\langle m\rangle\ln\bar{\Lambda}\frac{\tilde{\Psi}(v_M)}{v_M^3}\vec{v}_M, \label{Ciotti10},
\end{align}
where $\ln\bar{\Lambda}$ is a velocity averaged Coulomb logarithm, $\langle m\rangle$ is the mean mass of the background particles, and $\tilde{\Psi}(v_M)$ is the total fractional velocity volume function\footnote{We note that \cite{2010AIPC.1242..117C} dubs with velocity volume function $\Xi=4\pi\Psi$.} of the system. For the case of interest of a binary mass spectrum (i.e. stars and DM particles), we have
\begin{equation}
\tilde{\Psi}(v_M)=\frac{\mathcal{M}H_1+H_2}{\mathcal{M}+1},
\end{equation}
where $\mathcal{M}=M/\langle m\rangle$ and
\begin{align}
H_1(v_M)&=\frac{\Psi_*(v_M) +xy \Psi_{\rm DM}(v_M)}{1+xy};\cr H_2(v_M)&=(1+x)\frac{\Psi_*(v_M) +xy^2 \Psi_{\rm DM}(v_M)}{(1+xy)^2}.
\end{align}
In the expressions above, $x=n_{\rm DM}/n_*$ and  $y=m_{\rm DM}/m_*$ are the ratios of the number densities and masses of stars and DM particles, respectively and the $\Psi_{\rm DM}$ and $\Psi_{*}$ their fractional velocity volume functions. Stars and DM are not at energy equipartition, as the thermalization time scale would exceed the age of the universe of several order of magnitude due to the small mass of DM particles. Therefore, both species have the same velocity distribution dictated by their collective potential, hence $\Psi_{\rm DM}=\Psi_{*}=\Psi_{\rm DM}$.
As the DM is constituted by unknown elementary particles, with typical masses of the order of a few GeV, we have $m_{\rm DM}\ll m_*$ and $n_{\rm DM}\gg n_*$. The average mass $\langle m\rangle$ is expressed as:\ 
\begin{equation}
    \frac{1+xy}{x}m_*\ll m_*
,\end{equation}%
while
\begin{align}
    H_1(v_M)=\Psi(v_M); \quad H_2(v_M)=\frac{x}{(1+xy)^2}\Psi(v_M),
\end{align}
so that the total fractional velocity volume functions becomes:
\begin{align}
    \tilde{\Psi}(v_M)=H_1 +\frac{\langle m\rangle}{M}H_2=\Psi(v_M) +\frac{m_*}{M}\frac{1}{1+xy}\Psi(v_M).
\end{align}
For systems where the test mass $M\gg m_*$ the latter further simplifies and $\tilde{\Psi}(v_M)={\Psi}(v_M)$ and the DF formula can be simplified to:\ 
\begin{align}
    \frac{d\mathbf{v}_M}{dt}&=-16\pi^2G^2\rho_{\rm Tot}M\ln\bar{\Lambda}\frac{\Psi(v_M)}{v_M^3}\vec{v}_M.\label{simpleDM}
\end{align}
In practice, the presence of a DM distribution has enhances the DF of a factor, $\rho_{\rm Tot}/\rho_*$, with respect to the baryonic system with the same phase-space distribution. We note that \cite{2006MNRAS.370.1829S} also derived a simplified two-component DF formula for the cases of circular orbits in DM plus baryons \cite{1962AJ.....67..471K} cores, concluding that DF is reduced by the presence of DM of a factor $(1+\mathcal{R})^{-1}$ with respect to the case where the (same) mass is in one single component, following the light distribution.
\subsection{QuMOND case}\label{2.2}
We first present a naive general expression of the MONDian DF, making use of the fact that in QuMOND, we have:\ 
\begin{equation}
    F=\nu\left(\frac{Gm}{a_0 b^2}\right)\frac{GMm}{b^2}.
\end{equation}
For each particle of the background system, we defined the so-called MOND radius as
\begin{equation}
r_{\rm M}=\sqrt{\frac{Gm}{a_0}}.
\end{equation}
With the same arguments of Sect. \ref{newtonian}, we now find that the momentum variation of $M$ is: 
\begin{equation}
\Delta p=\nu(r_{\rm M}^2/b^2)GMm/v_M b. 
\end{equation}
The form of the collision rate factor remains unchanged, as it is a geometric quantity. The diffusion coefficient $D_v$ retains the same definition as in Eq. (\ref{Newt Dv}), but with a different integral over the impact parameter: 
\begin{equation}
I_b=\int_{b_{\rm min}}^{b_{\rm max}}\nu(r_{\rm M}^2/b^2)^2 db/b. 
\end{equation}
After substituting the expression in the DF formula, we obtain: 
\begin{equation}
\label{MOND dyn fric}
    \frac{d\mathbf{v}_M}{dt}=-16\pi^2I_bG^2\rho(M+m)\frac{\Psi(v_M)}{v_M^3}\vec{v}_M.
\end{equation}%
From Eq. (\ref{MOND dyn fric}) above, we notice immediately that the QuMOND DF is always larger than its Newtonian equivalent for the same baryonic mass distribution. This happens because $\nu^2>1$, thus returning a larger factor within the integral over the impact parameters. In addition, this is also in agreement with the fact that in MOND theories the far field behaviour of a given mass distribution falls off less rapidly than in Newtonian gravity, as implied by the logarithmic trend of the MOND potential.\\
\indent We now evaluate Eq. (\ref{MOND dyn fric}) for a simple choice of $\nu$. Knowing that $\nu$ has the asymptotic trends given in Eq. (\ref{Qu limits}) we start with a simplified form:
\begin{equation}
\label{simple nu}
    \nu(y)=1+1/\sqrt{y},
\end{equation}%
which is essentially the sum of the Newtonian and the deep MOND r\'egimes. In terms of $\mu$, Eq. (\ref{simple nu}) corresponds to $\mu(x)=1+(1-\sqrt{1+4x})/2x$. For such a choice, the correction factor is $\nu(r_{\rm M}^2/b^2)=1+b/r_{\rm M}$, and the integral on the impact parameters becomes:%
\begin{align}
   I_b&=\ln\frac{b_{\rm max}}{b_{\rm min}}+2\frac{b_{\rm max}-b_{\rm min}}{r_{\rm M}}+\frac{b_{\rm max}^2-b_{\rm min}^2}{2r_{\rm M}^2}= \cr
    &\cong \ln\Lambda+2\frac{b_{\rm max}}{r_{\rm M}}+\frac{b_{\rm max}^2}{2r_{\rm M}^2};
\end{align}%
where we assumed $b_{\rm min}\ll r_{\rm M}$ in the last equality. We note that the maximum impact parameter $b_{\rm max}$ in the context of Newtonian gravity is usually assumed to be some (arbitrary) scale distance\footnote{Such a cut-off length in neutral plasma physics is unequivocally constrained by the Debye length, see \cite{1965pfig.book.....S}} of the system at hand. In the original derivation of \cite{1943ApJ....97..255C}, where the system is infinite, the definition of $b_{\rm max}$ remains somewhat unclear. Some authors (see e.g. \citealt{VanAlbada} and references therein) typically take $b_{\rm max}$ of the order of the average inter-particle distance. Here, we follow the same assumption, noting that for a star of a solar mass, the MOND radius $r_{\rm M}$ is roughly $10^3$ AU; this is much smaller than the typical scale radius of a galaxy, but comparable to the mean inter-particle distance among stars. This established, the QuMOND DF then becomes
\begin{align}
\label{DF simple}
    \left(\frac{d\mathbf{v}_M}{dt}\right)\approx&-16\pi^2\left(\ln\Lambda+\frac{b_{\rm max}^2}{2r_{\rm M}^2}+2\frac{b_{\rm max}}{r_{\rm M}}\right)\times\cr
    &\times G^2\rho(M+m)\frac{\Psi(v_M)}{v_M^3}\vec{v}_M.
\end{align}
In practice, at first order in QuMOND, each particle behaves as a point-like source exerting the usual Newtonian $1/r$ potential plus an infinitely extended 'phantom DM' halo whose contribution depends on the specific form of $\nu$. Following this approach, the calculations as shown in Appendix \ref{phantom} leads to the same expression of Eq. (\ref{DF simple}). We note that Eq. (\ref{DF simple}) differs from what we would obtain using the simple derivation of $t_{2b}$ in Sect. 2 of \citealt{2004MNRAS.351..285C}; namely,  this derivation is proportional to the crossing time $t_{\rm cross}=1/\sqrt{G\rho}$ to evaluate the dynamical friction time scale with the \cite{Spitzer} relation $t_{\rm DF}=2t_{2b}m/(m+M)$. This is because their expression of $t_{2b}$ makes use of the force (\ref{f12}), implicitly assuming  the dMOND regime.\\
\indent Unlike the usual MOND interpolating function $\mu(x)$, the form of the QuMOND function $\nu(y)$ affects the behaviour of the MONDian force field. Adopting the commonly used Eq. (\ref{usual nu}) (see \citealt{2022MNRAS.517.3613K}), the DF expression becomes:
\begin{align} \label{DF usual}
    \frac{d\mathbf{v}_M}{dt}\approx& -16\pi^2\left(\ln\Lambda+\frac{b_{\rm max}^2}{2r_{\rm M}^2}+\frac{b_{\rm max}}{r_{\rm M}}-\frac{1}{2}\ln\frac{b_{\rm max}}{r_{\rm M}}-\frac{2+\ln2}{4}\right)\times \cr
    &\times G^2 \rho(M+m)\frac{\Psi(v_M)}{v_M^3}\vec{v}_M.
\end{align}%
The explicit derivation is given in Appendix \ref{usual calc}. \\
\indent Noteworthy, we observe in both cases that the MOND correction with respect to the Newtonian baryon only case is proportional to a function of the only $b_{\rm max}/r_{\rm M}$
\begin{align}
    \left(\frac{d\mathbf{v}_M}{dt}\right)_{\rm QuMOND}&-\left(\frac{d\mathbf{v}_M}{dt}\right)_{\rm Newt}\propto \frac{b_{\rm max}^2}{2r_{\rm M}^2}+2\frac{b_{\rm max}}{r_{\rm M}} \quad , \cr
    &\frac{b_{\rm max}^2}{2r_{\rm M}^2}+\frac{b_{\rm max}}{r_{\rm M}}-\frac{1}{2}\ln\frac{b_{\rm max}}{r_{\rm M}}-\frac{1}{2}.
\end{align}%
The same term $b_{\rm max}^2/2r_{\rm M}^2$ always dominates the MOND correction of DF, regardless of the exact form of $\nu$, while the smaller terms are affected by the particular choice.
Recently, a number of authors (see e.g. \citealt{2017ApJ...836..152L}) proposed an expression for $\nu$ supported by observational data proportional to the form of radial acceleration relation. However, its functional form is rather complex thus making the explicit evaluation of Eq. (\ref{MOND dyn fric}) particularly cumbersome.\\ 
%%%%%%%%%%%%%%%%%%%%%%%%%%%%%%%%%%%%%%%
\begin{figure}
        \centering
        \includegraphics[width=\columnwidth]{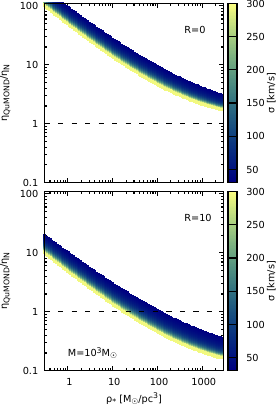}
\caption{Ratio of the QuMOND and Newtonian DF coefficients as function of the stellar density, for $\mathcal{R}=0$ (top) and 10 (bottom), for the case of a $M=10^3M_{\odot}$ test mass. The dependence on the local velocity dispersion $\sigma$ is colour coded.}
\label{fig1}
\end{figure}
%%%%%%%%%%%%%%%%%%%%%%%%%%%%%%%%%%%%%%%%%%%%%%%%%%%%%%%%%%%%%%%%
\indent As an example, in the top panel of Figure \ref{fig1} we show for a test mass $M=10^3M_{\odot}$ the ratio $\eta_{\rm QuMOND}/\eta_{N}$ of the QuMOND and Newtonian dynamical friction coefficient as a function of the stellar density, $\rho_*$, for different values of the velocity dispersion $\sigma$ (see e.g. \citealt{binney}) indicated in the figure with the colour map. In the corrective QuMOND term, $X=b_{\rm max}^2/2r_M^2+b_{\rm max}/r_M-1/2\ln(b_{\rm max}/r_M)-(2+\ln 2)/4$ of Eq. (\ref{DF usual}) and in the standard Coulomb logarithm, we assumed $b_{\rm max}=(\rho/M_{\odot})^{-1/3}$ and $b_{\rm min}=G(M+m)/\sigma^2$. It is evident that for an isolated baryon-only system (corresponding to a dark to luminous matter ratio, $\mathcal{R}=0$) the DF in QuMOND is augmented (up to a factor $\approx 10^2$) at low density, comparable to the typical intra-galactic density of $10^{-1}$ stars per pc$^3$, with respect to the Newtonian case. Vice-versa (and as expected) in denser systems/regions the QuMOND correction is negligible, in particular for large values of the velocity dispersion, $\sigma$. Of course, a fair comparison between Newtonian and MONDian DF has to be made considering the presence of DM. In the bottom panel of Figure \ref{fig1} we show $\eta_{\rm QuMOND}/\eta_{N}$ for the same values of $\rho_*$ and $\sigma$ and $\mathcal{R}=10$ (a typical DM to baryonic matter ratio). In this case, the Newtonian DF is given by Eq. (\ref{simpleDM}). Remarkably, in this case the QuMOND DF is substantially lower than the Newtonian one including the dark mass contribution (substantially confirming the results of \citealt{2006MNRAS.370.1829S} for a simplified circular orbit model and of \citealt{2008MNRAS.386.2194N,2021MNRAS.503.2833R} for the dynamics of bars in spiral galaxies), for almost all astrophysically relevant combinations of stellar density and velocity dispersion. In the low $\rho_*$ regime (say below $\approx 8M_\odot{\rm pc}^{-3}$), the simplified QuMOND DF expression dominates over its Newtonian counterpart. This leads us to speculate that DM-dominated low $\rho_*$ systems, such as ultra-faint dwarf galaxies, could be used as a testing ground for MOND theories versus $\Lambda$CDM using the different behaviour of the DF (see e.g. \citealt{2021A&A...653A.170B,universe10030143}). We stress the fact that in finite systems, both in Newtonian gravity and MOND, the orbital structure (i.e. the velocity anisotropy profile of the background particle distribution) has a relevant effect on the magnitude of DF. For example, a test mass, $M,$ moving on a highly radial orbit in a radially anisotropic spherical system suffers less DF than in a systems with isotropic velocity distribution. In this case, in MOND, where only stars contribute to DF, the drag is further reduced when $M$ reaches radii where no mass is present. A similar argument has been applied (among others) by \cite{10.1111/j.1365-2966.2009.14745.x} to explain in MOND the survival of the globular cluster systems of the Fornax dwarf spheroidal.
\section{Mean field approach} \label{S3}
A more rigorous evaluation of the DF expression can be carried out using the mean field formalism developed by \cite{1983Ap&SS..97..435K} (see also \citealt{1968ApJ...152.1043G} and \citealt{1980PhR....63....1K}) in the context of Newtonian gravity that can be applied to any kind of long-range force, not necessarily obeying the superposition principle.\\
\indent We consider the usual system of $N$ equal mass $m$ (field) particles, with coordinates $\mathbf{r}_i$ and momenta, $\mathbf{p}_i=m\mathbf{v}_i$, described by the (time-dependent) phase-space distribution function,\footnote{Note: $\mathcal{F}$ is not the one particle distribution function $f(r,v)$ entering the collisionless Boltzmann equation, but rather the phase-space distribution of a $6N$ degrees of freedom Hamiltonian system entering the Liouville equation.} $\mathcal{F}(\mathbf{r}, \mathbf{p}; t),$  a test particle $M$, with a coordinate, $\mathbf{r}_0=\mathbf{R}$, and momentum, $\mathbf{p}_0=\mathbf{P}=M\mathbf{v}_M$, which  perturbs the initial phase-space distribution $\mathcal{F}_0$. By virtue of the third law of dynamics, such perturbation corresponds to the force decelerating the test mass, $M$. We assume the field particles to be statistically uncorrelated in their initial state with Maxwellian velocity distribution. Moreover, we also take the limit of infinite, homogeneous distribution of field particles. In his original work, \cite{1983Ap&SS..97..435K} initially assumed an external potential, $\Phi$, confining the system (see Appendix \ref{Kandrup calc} below). We note that MOND systems with self-consistent gravitational field, $g_{\rm in}$, embedded in an external gravitational field $g_{\rm ext}$ are prone to the so-called external field effect (hereafter EFE). The latter implies that for $g_{\rm in}<a_0<g_{\rm ext}$ the system is purely Newtonian, while for $g_{\rm in}<g_{\rm ext}<a_0$ the system behaves as a Newtonian model with rescaled gravitational constant, $G^\prime=Ga_0/g_{\rm ext}$.\\
\indent Under the assumptions given above, as usual, the average number density of particles, $n=N/{\rm Vol}$, should be taken constant as $N, {\rm Vol}\rightarrow\infty$. The dynamical friction force $F_0^{\rm fr}$ can be therefore obtained simply as $\langle F_0^{\rm tot}\rangle_{\mathcal{F}}$.\\
\indent In the same fashion of the special relativistic extensions of DF (see e.g. \citealt{1994MNRAS.270..205S} and \citealt{2023A&A...677A.140C}), we shift to the frame of reference of $M$ such that for the new coordinates, $\hat{\mathbf{r}}, \hat{\mathbf{p}}$, the linear evolution operator for $\mathcal{F}$ takes the form $\dot{\mathcal{F}}(\hat{\mathbf{r}}_i, \hat{\mathbf{p}}_i; t)=-i\mathcal{L}[\mathcal{F}]$. See Eqs. (\ref{evol1}) and (\ref{evol2}) in the appendix below for its explicit expression.\\
\indent The linear formulation of the evolution equation is allowed since MOND (and therefore QuMOND), although non-linear, is still a Lagrangian and local\footnote{MOND is local in the sense that the density at point $\xi$ equals a differential operator acting on the potential in $\xi$. Vice versa, for the case of $1/r^{\alpha}$ forces  with $\alpha\neq 2$, obeying the superposition principle this is not true (\citealt{8bbc344b-39d3-31c4-89fe-4b604b1a5849}). In practice, for such general long-range interactions, density and potential are related only via integral relation over the whole domain occupied by the system (\citealt{2011IJBC...21.2279D,2013MNRAS.431.3177D}).} theory. At this stage, it is not necessary to consider explicitly the \cite{1984ApJ...286....7B} (or \citealt{2010MNRAS.403..886M}) Langrangian, since it is enough to consider the particle's energies.\\
\indent The significant difference with respect to the Newtonian case explored in \cite{1983Ap&SS..97..435K} is that in MOND, we cannot substitute the potentials in a form: 
\begin{equation}
W_i=\sum_{j\neq i}W_{ij}(|\mathbf{r}_i-\mathbf{r}_j|), 
\end{equation}
due to the non-linearity of MOND implied by the absence of a Superposition principle. Such potentials however can be rather calculated in the QuMOND formalism (\ref{QuMOND}).\\
\indent A general expression for the DF force on the test particle can be written formally as
\begin{align}\label{formaldfqumond}
    \mathbf{F}_0^{\rm fr}=-\beta \int d\Sigma \mathcal{F}_0\left(\sum_{j=1}^N \mathbf{F}_i^{\rm tot}\right) \int_0^t d\tau G_{\mathcal{L}}(\tau\rightarrow t)\left[\sum_{i=1}^N \mathbf{v}_i\cdot\mathbf{F}_i^{\rm tot}\right],
\end{align}%
where $G_{\mathcal{L}}$ is the Greenian of the operator $\mathcal{L}$ and $d\Sigma=d^{3N}\mathbf{r}d^{3N}\mathbf{p}$ is the differential element in phase space. The details of the derivation are discussed in Appendix \ref{Kandrup calc}.\\
\indent We now sketch the main simplifying hypotheses needed to perform practically the integrals in Eq. (\ref{formaldfqumond}). Assuming that the system under consideration has a finite memory (as implied by its ergodicity), we can replace the time integration in $\tau$ on the finite interval $[0; t]$ with an integration extended over the semi-infinite interval $[0; \infty)$. Moreover, making use of the standard linear trajectory approximation (see e.g. \citealt{1977lost.book.....T}, see also \citealt{1994MNRAS.270..205S}), allows us to simplify $G_{\mathcal{L}}(\tau\rightarrow t)[Q]\cong Q(t-\tau)$. The latter is the most delicate step of the present calculation, since in this approximation, the effects of the different field particles decouple that is, in principle, not valid in non-linear theory.\\
\indent Under the assumptions listed above, the DF formula becomes:
\begin{align}
\label{Kandrup MOND}
    \frac{d\mathbf{v}_M}{dt}=\frac{\mathbf{F}_0^{\rm fr}}{M}\cong&-\frac{3G^2M \rho|_{v<V_M}}{\langle v^2\rangle} \int_0^\infty\iint_\Omega f(\mathbf{v})\frac{\mathbf{v}\cdot(\mathbf{s}-\tilde{\mathbf{v}}\tau)}{|\mathbf{s}-\tilde{\mathbf{v}}\tau|^3}\cr
    &\times \nu\left(\frac{r_{\rm M}^2}{s^2}\right) \nu\left(\frac{r_{\rm M}^2}{|\mathbf{s}-\tilde{\mathbf{v}}\tau|^2}\right)\frac{\mathbf{s}}{s^3}d^3\mathbf{v}d^3\mathbf{s}d\tau,
\end{align}%
%%%%%%%%%%%%%%%%%%%%%%%%%%%%%%%%%%%%%%%
\begin{figure}
        \centering
        \includegraphics[width=\columnwidth]{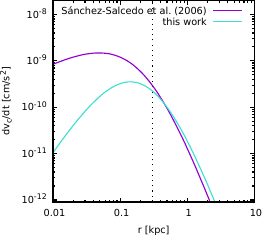}
\caption{Radial profiles of the DF force (per unit mass) in the circular orbit approximation as given by Eqs. (\ref{sanchez},\ref{thiswork}); purple and cyan lines, respectively.}
\label{fig2}
\end{figure}
%%%%%%%%%%%%%%%%%%%%%%%%%%%%%%%%%%%%%%%%%%%%%%%%%%%%%%%%%%%%%%%%
where $\tilde{\mathbf{v}}=\mathbf{v}-\mathbf{v}_M$, $\Omega$ is the phase space volume occupied by the system and only the field particles slower than $V_M$ induce the drag, so that we only for account their density, $\rho|_{v<V_M}$. At this stage, the problem is reduced to solving a three dimensional integral for a given choice of the QuMOND interpolation function, $\nu$.\\
\indent Eq. (\ref{Kandrup MOND}) in the Newtonian limit (i.e. $\nu\rightarrow 1$) can be solved by an integration, incrementally applying:
\begin{equation}
    \nabla\cdot\frac{\mathbf{s}}{s^3}=-4\pi\delta^3(\mathbf{s})
\end{equation}
and
\begin{equation}
    \mathbf{v}\cdot\frac{\mathbf{s}-\tilde{\mathbf{v}}\tau}{|\mathbf{s}-\tilde{\mathbf{v}}\tau|^3}=\nabla\cdot\frac{\mathbf{v}}{|\mathbf{s}-\tilde{\mathbf{v}}\tau|},
\end{equation}
thus, we have:\ 
\begin{equation}
    \int\left(\mathbf{v}\cdot\frac{\mathbf{s}-\tilde{\mathbf{v}}\tau}{|\mathbf{s}-\tilde{\mathbf{v}}\tau|^3}\right)\frac{\mathbf{s}}{s^3}d^3\mathbf{s}=4\pi\frac{\mathbf{v}}{\tilde{v}\tau}.
\end{equation}%
Here, the term $\tilde{v}\tau=b$ is the standard impact parameter (see again \citealt{binney}), thus, the integral
\begin{equation}
    \int_0^{\infty}\frac{d\tau}{\tau}=\int_0^{\infty}\frac{db}{b}
\end{equation}
returns the usual Coulomb logarithm (see e.g. \citealt{1972ASSL...31...13K}), while the integral in $d^3\mathbf{v}$ is rewritten as in the Newtonian case, therefore yielding Eq. (\ref{Chandra}).\\
\indent The QuMOND case is, even for the simple form of $\nu$ given in Eq. (\ref{simple nu}), plagued by its intrinsic complexity. Analogously to the naive formula in Eq. (\ref{MOND dyn fric}) we notice the presence of two factors of $\nu$. In this mean field formalism, the arguments for the two functions $\nu$ in (\ref{Kandrup MOND}) are still slightly different. From such difference $-\tau\tilde{\mathbf{v}}$, we recovers the logarithmic term in $\tau$, as well as any additional QuMOND term. At variance with the Newtonian case, which requires some cut-off value only for the integral over $d\tau$, in the QuMOND DF formula of Eq. (\ref{Kandrup MOND}), the integral over $d^3\mathbf{s}$ is also diverging; therefore, it requires a cut-off at $b_{\rm max}$. This yields terms of the form $b_{\rm max}/r_{\rm M}$ and $b_{\rm max}^2/r_{\rm M}^2$, which are vastly in agreement with the simplified approach described in Sect. \ref{2.2}.\\
\indent In the deep MOND regime, where $\nu(y)\rightarrow y^{-1/2}$, we can simplify Eq. (\ref{Kandrup MOND}) as:
\begin{align}
\label{Kandrup dMOND}
    &\frac{d\mathbf{v}_M}{dt}=-\frac{3G^2M\rho_*|_{v<V_M}}{\langle v^2\rangle r_{\rm M}^2} \int_0^\infty\iint_\Omega f(\mathbf{v})\frac{\mathbf{v}\cdot(\mathbf{s}-\tilde{\mathbf{v}}\tau)}{|\mathbf{s}-\tilde{\mathbf{v}}\tau|^2} \frac{\mathbf{s}}{s^2}d^3\mathbf{v}d^3\mathbf{s}d\tau=\cr
    &=-\frac{12\pi^2 G^2 M\rho_* b_{max}^2}{\langle v^2\rangle r_M^2}\Psi(V_M)\int_{v<V_M} \frac{f(v)}{\tilde{v}} \left[\frac{\pi^2}{8}\mathbf{v} -\left(1+\frac{3}{16}\pi^2\right)\frac{\mathbf{v}\cdot\tilde{\mathbf{v}}}{\tilde{v}^2}\tilde{\mathbf{v}}\right]d^3\textbf{v}.
\end{align}%
%%%%%%%%%%%%%%%%%%%%%%%%%%%%%%%%%%%%%%%%%%%%%%%5
We recall that \cite{2006MNRAS.370.1829S} (see also the numerical simulations of \citealt{2021A&A...653A.170B}), used the circular orbit approximation and the estimates from \cite{2004MNRAS.351..285C} to formulate a local dMOND DF expression for a spherical system where the radial friction on $M$ is expressed as:
\begin{equation}\label{sanchez}
\frac{dv_c}{dt}=-\frac{16\pi^2\log\Lambda G^2M\rho_*(r)}{v_c^2}\left(\frac{a_0^2}{\sqrt{2}a_{\rm typ}^2}\right)\Psi(v_c).
\end{equation}
In the expression above, $a_{\rm typ}\approx GM_*/r_s^2$ is the typical Newtonian acceleration of the system with scale radius $r_s$, and the circular velocity profile is given (see \citealt{1983ApJ...270..365M}) as a function of the stellar radial mass profile $M_*(r),$ as follows:
\begin{equation}
v_c(r)=\left[Ga_0M_*(r)\right]^{1/4}.
\end{equation}
Taking the same assumptions applied in \citealt{2006MNRAS.370.1829S}, we can  simplify Eq. (\ref{Kandrup dMOND}) as $\mathbf{v}$ and $\mathbf{s}$ are generally likely to be perpendicular, while $\mathbf{v}$ and $\tilde{\mathbf{v}}$ are almost parallel. This allows us to then simplify the expression of the term $\mathbf{v}\cdot(\mathbf{s}-\tilde{\mathbf{v}}\tau)/|\mathbf{s}-\tilde{\mathbf{v}}\tau|^2$  in scalar form and perform the integral in $d\tau$ before those in velocity, $dv$, and in space, $ds$, having imposed spherical symmetry. Standard (albeit tedious) algebra then yields
\begin{equation}\label{thiswork}
\frac{dv_c}{dt}=-\frac{12\pi^2G^2M\rho_*(r)b_{\rm max}^2\log 2}{\langle v^2\rangle r_M^2}\Psi(v_c),
\end{equation}
where the dMOND mean square velocity is given (\citealt{1995ApJ...455..439M}) as 
\begin{equation}\label{sigmadmond}
\sigma^2=\langle v^2\rangle=\sqrt{4Ga_0M_{*}/81},
\end{equation}
for a system with total baryonic mass $M_{*}$ and $b_{\rm max}$ (typically of the order of the local mean inter-particle distance) is fixed to a scale radius $r_s$.\\
\indent In Fig. \ref{fig2}, we show the DF acceleration in the limit of circular orbits on a mass  of $M=2\times 10^5M_\odot$, as function of the radial distance, $r$, as given in Eqs. (\ref{sanchez}) and (\ref{thiswork}), sinking in a cored spherical model (cfr. also Eq. \ref{gamma0} in the next section) with $M_*=2\times 10^7M_\odot$ and a core radius of $r_s=0.3$ kpc, corresponding to the model of a Fornax-like dwarf galaxy considered by \citealt{2006MNRAS.370.1829S}. With such a choice of parameters, the constant dMOND velocity dispersion given by Eq. (\ref{sigmadmond}) is $\approx 11{\rm km}/{\rm s}$. We note that  at large radii (e.g. above the core radius indicated in Figure by the vertical dashed line), the two expressions are somewhat comparable; however, in the inner part of the system, the DF is considerably lower in our approximated formulation. This is because for $r\rightarrow 0$, the l.h.s. of Eq. (\ref{thiswork}) is mostly dominated by the contribution of $\Psi(v_c)$, that for the specific (but questionable) choice of a Maxwellian velocity distribution is expressed as:
\begin{equation}
\Psi(v_c)=(4\pi)^{-1}\left[{\rm erf}\left(\frac{v_c}{\sqrt{2}\sigma}\right)-\sqrt{\frac{2}{\pi}}\frac{v_c}{\sigma}\exp\left(-\frac{v_c^2}{2\sigma^2}\right)\right],
\end{equation}
then falls rapidly to zero, while in Eq. (\ref{sanchez}), it is partially compensated by the factor $1/v_c^2$.\\
\indent Interestingly, the QuMOND $N-$body simulations of \cite{2021A&A...653A.170B} offer a hint that Eq. (\ref{sanchez}) overestimates the DF in the inner region of cored systems; whereas at large radii, the semianalytical estimate of \cite{2006MNRAS.370.1829S} is well matched by the simulations. The discrepancy is interpreted as an effect of the so-called dynamical buoyancy or core stalling (\citealt{2021MNRAS.502.1441B}) in a nearly harmonic core, preventing the further sinking of the test particle, $M$. We note that in MOND (and in particular in the dMOND limit assumed here), an almost constant density core does not exert a harmonic potential, thereby potentially undermining  the core-stalling argument in MOND simulations. This leads to conjecture that the intrinsically less effective dMOND DF in the $N-$body simulation is indeed responsible for the discrepancy with the prediction of Eq. (\ref{sanchez}).
\section{Numerical experiments}
\begin{figure}
        \centering
        \includegraphics[width=\columnwidth]{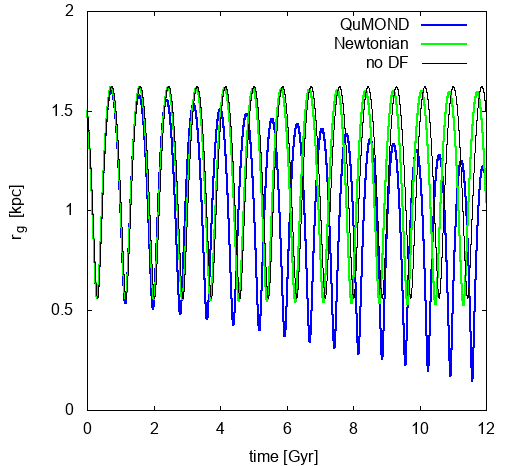}
\caption{Evolution of the galactocentric radius, $r_g$, for a star cluster orbiting through a dwarf galaxy in Newtonian gravity (green curve) and QuMOND (blue curve). The thin black line marks the case without dynamical friction.}
\label{fig3}
\end{figure}

To clarify the different behaviour of DF in the QuMOND and Newtonian gravity model of a given stellar system, it is useful to perform a simple numerical experiment in a set-up where Newtonian gravity with DM the friction on the test object, $M,$ would be negligible, while the gravitational field of the parent system would be in the MONDian regime, $g\ll a_0$, over a broad interval of radii.\\
\indent We consider a $M=10^2M_\odot$ star cluster moving through a dwarf galaxy with stellar mass of $M_*=2\times 10^5M_{\odot}$ and scale radius of $r_s=0.8$ kpc (parameters compatible to those of the Draco ultra-faint dwarf galaxy, \citealt{2006ApJ...640..252M}), with a stellar distribution given by
\begin{equation}\label{gamma0}
\rho_*(r)=\frac{3}{4\pi}\frac{M_{*}r_{s}}{(r+r_{s})^{4}},
\end{equation}
corresponding to a \cite{1993MNRAS.265..250D} $\gamma-$model for $\gamma=0$. The associated QuMOND gravitational potential given by Eq. (\ref{QuMOND}) is (by construction) identical to the potential exerted by a Newtonian system with stellar density, as in Eq. (\ref{gamma0}), plus a DM halo with $\rho_{\rm DM}=\tilde{\rho}-\rho_*$ (cfr. Eq. \ref{rhoaug}). By contrast, the DF force experienced by $M$ is different in the two paradigms:\ Newtonian and MOND, with Eq. (\ref{simpleDM}) displaying the first case\footnote{We recall that in a multimass system (e.g. stars and elementary particle sized DM), when the test mass, $M,$ is much larger than the mean mass, $\langle m\rangle, $ then the prefactor of Eq. (\ref{Chandra}) is dominated by $M$ times the total mass density, as described in e.g. \cite{2021isd..book.....C}.} and Eq. (\ref{DF usual})  the second.\\ 
\indent Using the numerical approach discussed in Appendix \ref{nummeth}, (see also \cite{2020A&A...640A..79P,2020IAUS..351...93D,2022IAUS..364..152D}), we integrated different orbits under the effect of the same gravitational field and the two different DF expressions, under the assumption of local Maxwellian approximation (i.e. at all radii the velocity distribution is an isotropic Maxwell-Boltzmann with velocity dispersion approximated by $\sigma=\sqrt{2||\Phi||}$). In Figure \ref{fig3}, we show the evolution of the galactocentric distance, $r_g$, of $M$ over 12 Gyr for an orbit of initial ellipticity, $e=0.49$. The green and blue curves mark the Newtonian and QuMOND simulations, respectively, while the black curve is the unperturbed orbit in the static potential of the model. We observe that while the Newtonian DF only minimally alters the orbit of $M$ at around 9 Gyr, in QuMOND, the test particle suffers a rather strong orbital decay, with the pergalactic radius falling from $\approx 1.63$ kpc down to $\approx 1.2$ kpc at 12 Gyr. This is remarkably in agreement with the analytical estimates of \cite{2004MNRAS.351..285C}, who predicted a MOND inspiral time of less than a Hubble
time for an object that is considerably more massive than a star to the central regions of a Draco-like system. This time span is much shorter than its Newtonian analog and argued that this could be the reason behind the absence of star clusters in similar dwarf galaxies. 
%%%%%%%%%%%%%%%%%%%%%%%%%%%%%%%%%%%%%%%
%%%%%%%%%%%%%%%%%%%%%%%%%%%%%%%%%%%%%%%%%%%%%%%%%%%%%%%%%%%%%%%%
\section{Conclusions}
We investigated the dynamical friction in the Quasi linear formulation of MOND. Using a simple dimensional analysis, we find that the expression for the MONDian DF is augmented with respect to its Newtonian counterpart of a factor proportional to the MOND radius of the star, $r_{\rm M}$. This additional term becomes relevant when the maximum impact parameter (proportional to the average interparticle distance, \citealt{Spitzer}) is $b_{\rm max}\gg r_{\rm M}$, as seen by Eqs. (\ref{DF simple} and \ref{DF usual}). In practice, in a dense star cluster or in galactic nuclei with a mean stellar density and where the typical inter-stellar distance is always smaller than $r_{\rm M}$, the DF is always purely Newtonian. Vice versa, in low-density systems, such as ultra-faint dwarf galaxies (\citealt{2019ARA&A..57..375S}) the $b_{\rm max}/r_{\rm M}$ correction is always of the order of the Coulomb logarithm, thus enhancing the (stellar) DF for a system (where, in principle, it should be negligible). In addition, we find that the explicit form of the QuMOND correction is dependent on the specific choice of the interpolating function $\nu$. However, the strength of the enhancing term does not vary significantly for such different forms of $\nu$. We also compared our results with a Newtonian plus DM expression obtained from the \cite{2010AIPC.1242..117C} formalism for DF in presence of a mass spectrum, confirming that in almost all astrophysically relevant regimes, the Newtonian DF (in presence of dark matter) dominates over its MOND counterpart.\\
\indent Moreover, we have extended the mean field treatment of DF, pioneered by \cite{1983Ap&SS..97..435K} in the Newtonian case to QuMOND. In this framework, alternatively to the fluctuation-based approach by \cite{2004MNRAS.351..285C}, we recovered an integral expression that contains the usual dependence on $\nu$. Unfortunately, the explicit evaluation of the DF force becomes rapidly cumbersome (at variance with the simpler Newtonian case) even in the dMOND limit; thus, we are forced to integrate Eq. (\ref{Kandrup MOND}) numerically (specific cases of this will be discussed in a future work).  Working out in QuMOND the two-body relaxation time (and therefore the DF coefficient of Eq. \ref{tmond} via the relation $t_{DF}=2t_{2b}m/M$) with the fluctuation approach in Fourier space of \cite{2004MNRAS.351..285C} may, in principle, be possible; however, it would imply a double application of the classical Poisson equation and a single non-linear algebraic step is hindered by the implicit relation between the Fourier transform of the QuMOND potential and the associated auxiliary density $\tilde{\rho}$. Using the mean field approach, we worked out an approximate formula, valid in the deep MOND regime for circular orbits analogous to the one presented in \cite{2006MNRAS.370.1829S}. For the case of a globular cluster orbiting in a isolated Fornax-like dwarf galaxy, the two expressions are comparable, at least for radii larger than the core radius of the dwarf, below which our expression underestimates the DF. This result appears to be in agreement with the $N-$body simulation of \cite{2021A&A...653A.170B}.\\
\indent Interestingly, simple numerical integration of a test particle in a QuMOND and its Newtonian equivalent system evidence that dynamical friction acts in a different way in the two paradigms, being considerably stronger in the first if the model is in a deep MOND regime (such as for DM-dominated dwarf galaxies). However, we stress the fact that in MOND, we cannot simply add a semi-analytic DF force to a mean field potential obtained solving numerically Eq. (\ref{MOND}) for a given density (either imposed extrapolated from particles position) in the same fashion as in \cite{2014ApJ...795..169A,2014ApJ...785...51A,2022IAUS..364..152D}.  Therefore, our simple numerical estimates could potentially overestimate the MONDian DF. Using the quasi-linear formulation of MOND could, in principle, allow us to incorporate the contribution of DF (and possibly density fluctuations on a scale smaller than the particle-resolution) by adding it before evaluating $\tilde{\rho}$ to solve Eq. (\ref{QuMOND}) (see \citealt{di_cintio_2023_10397120}). In conclusion, it is also important to recall that as a consequence of the EFE, in systems with $g_{\rm int}<g_{\rm ext}<a_0$, the dynamics is essentially Newtonian with a rescaled $G$. In those cases, the DF force would be given by Eq. (\ref{Chandra}) augmented by the multiplicative factor $(a_0/g_{\rm ext})^2$.  Therefore, a full mean field QuMOND treatment including the DF expression (discussed in Sect. 3) is only valid  for an isolated system.
\begin{acknowledgements}
  We express gratitude to Jan Pflamm-Altenburg, Pavel Kroupa and Luca Ciotti for the useful discussions at an early stage of this work. We also thank the anonymous Referee for his/her useful remarks that helped improving the presentation of our results.
\end{acknowledgements}
   \bibliographystyle{aa} % style aa.bst
   \bibliography{biblio} % your references Yourfile.bib
\begin{appendix}
%%%%%%%%%%%%%%%%%%%%%%%%%%%%%%%%%%%%%%%%%%%%%%%%%%%%%
\section{Effective dark matter halos}
\label{phantom}
Combining Eqs. (\ref{Newton}) and (\ref{QuMOND}), we evaluate the equivalent matter distribution from Eq. (\ref{rhoaug}) as%
\begin{equation}
    \tilde\rho=-\frac{1}{4\pi G}\vec{\nabla}\cdot\vec{g}=-\frac{1}{4\pi G}\vec{\nabla}\cdot\left[\nu(|\vec{g}_N|/a_0)\vec{g}_N\right].
\end{equation}%
For a point particle, the density becomes $\rho(\vec{r})=m\delta^3(\vec{r})$, so that $\vec{g}_N(\vec{r})=-{Gm}/{r^3}\vec{r}$, and%
\begin{equation}
    \tilde\rho=-\frac{1}{4\pi G}\vec{\nabla}\cdot\left[-\frac{Gm}{r^3}\vec{r}\nu\left(\frac{Gm}{a_0 r^2}\right)\right]=\frac{m}{4\pi}\vec{\nabla}\cdot\left[\frac{\vec{r}}{r^3}\nu\left(\frac{r_{\rm M}^2}{r^2}\right)\right]
.\end{equation}%
Substituting the simple interpolation (\ref{simple nu}), we then find%
\begin{equation}
    \tilde\rho=\frac{m}{4\pi}\vec{\nabla}\cdot\left(\frac{\vec{r}}{r^3}\right)+\frac{m}{4\pi r_{\rm M}}\vec{\nabla}\cdot\left(\frac{\vec{r}}{r^2}\right)=m\delta^3(\vec{r})+\frac{m}{4\pi r_{\rm M} r^2},
\end{equation}%
%The first divergence appearing in the equation above yields $4\pi\delta^3(\vec{r})$, as it is applied to its own Green function. To evaluate the second term, it is useful to express the divergence in spherical coordinates as
%\begin{equation}
%\vec{\nabla}\cdot(V_r\hat{r}+V_{\theta}\hat{\theta}+V_{\phi}\hat{\phi})=\frac{\partial(V_r r^2)}{r^2 \partial r}+\frac{\partial(V_{\theta}\sin\theta)}{r\sin\theta \partial\theta}+\frac{\partial V_{\phi}}{r\sin\theta \partial\phi}.
%\end{equation}%
%As the potential generated by a point particle is spherically symmetric, with the radial component $V_r=1/r$, we can rewrite the divergence as%
%\begin{equation}
%    \vec{\nabla}\cdot(r^{-1}\hat{r})=\frac{\partial(r^{-1}r^2)}{r^2 \partial r}=r^{-2}.
%\end{equation}%
%We can hence write down the equivalent matter distribution as%
%\begin{equation}
%    \tilde\rho(\vec{r})
%\end{equation}%
where the first term here is the real point particle, and the second one represents its phantom dark matter halo $\rho_{pDM}=\tilde\rho-\rho$. Its cumulative mass function is%
\begin{equation}
    m_{DM}(r)=\int \frac{m}{4\pi r_{\rm M} r^2}4\pi r^2 dr=m\frac{r}{r_{\rm M}}.
\end{equation}
Following the approach of Chandrasekhar, the impulsive force generated by the particle and its effective DM halo is then
\begin{equation}
    F=\frac{GMm(b)}{b^2}=\frac{GMm}{b^2}+\frac{GMm}{r_{\rm M} b}.
\end{equation}%
The associated momentum variation is $\Delta p=Fb/v_M={GMm}/{v_M}\left(b^{-1}+r^{-1}_{\rm M}\right)$ and then the diffusion coefficient becomes
\begin{align}
    D_v&=\int\Delta p^2 2\pi nv_M d db=2\pi G^2 M^2\frac{m^2 n}{v_M}\int\left(\frac{1}{b}+\frac{1}{r_{\rm M}}\right)^2 b db.
\end{align}%
Keeping in mind that $b_{m}\ll r_{\rm M}<b_{\rm max}$, the integral can be evaluated as $\approx \ln\Lambda +2b_{\rm max}/r_{\rm M}+b_{\rm max}^2/2r_{\rm M}^2$. The QuMOND DF force is then%
\begin{align}
    \frac{d\mathbf{v}_M}{dt}=-2\rho(M+m)D_v\frac{N(|\vec{v}|<v_M)}{v_M^2}\vec{v}_M= \cr
    =-16\pi^2\left(\ln\Lambda +2\frac{b_{\rm max}}{r_{\rm M}}+\frac{b_{\rm max}^2}{2r_{\rm M}^2}\right)G^2\rho(M+m)\frac{\vec{v}_M}{v_M^3}\int_0^{v_M}v^2f(v)dv. 
\end{align}

\section{Explicit choice of the QuMOND interpolating function}
\label{usual calc}
If the usual choice (\ref{usual nu}) of the QuMOND interpolating function $\nu$ is used in Eq. (\ref{MOND dyn fric}), we have
\begin{equation}
    \nu(r_{\rm M}^2/b^2)=\frac{1}{2}+\sqrt{\frac{1}{4}+\frac{b^2}{r_{\rm M}}} \Rightarrow \nu(b)^2=\frac{1}{2}+\frac{1}{2}\sqrt{1+4\frac{b^2}{r_{\rm M}^2}}+\frac{b^2}{r_{\rm M}^2}.
\end{equation}%
Unfortunately, here we cannot take the usual approximation for small or large impact parameters $b$, since the integral runs from $b_{\rm min}<r_{\rm M}$ to $b_{\rm max}>r_{\rm M}$. Vary large values of $b_{\rm max}$ (say of the order of the system size), return an important contribution already in the Newtonian case, in the form of large values of $\log\Lambda$. This must also hold true {a fortiori} in MOND, as in the far-field the forces decay as $1/r$ rather than $1/r^2$. Therefore, we cannot expand $\nu$ to low-order terms in $b/r_{\rm M}$.\\
\indent We must then solve the exact integral $\int \nu(b)^2 db/b$, where the square root term yields
\begin{align}\label{squareroot}
    \int\sqrt{1+4b^2/r_{\rm M}^2}\frac{db}{b}&=\int\frac{\sqrt{1+x^2}}{x}dx= \cr
    &=\sqrt{1+x^2}-\text{arctanh}(\sqrt{1+x^2})+{\rm const}.= \cr
    &=\sqrt{1+x^2}-\frac{1}{2}\ln\frac{\sqrt{x^2+1}+1}{\sqrt{x^2+1}-1}+\left({\rm const}-\frac{\pi}{2}i\right), \cr
\end{align}%
where $x=2b/r_{\rm M}$. Evaluating Eq. (\ref{squareroot}) for $x=2b_{\rm min}/r_{\rm M}\rightarrow 0$ returns approximately
\begin{align}
    (1+x^2/2)-\frac{1}{2}\ln\frac{1+1}{(1+x^2/2)-1}=\ln\frac{b_{\rm min}}{r_{\rm M}}+1,
\end{align}%
while for $x=2b_{\rm max}/r_{\rm M}\rightarrow\infty$ becomes asymptotically 
\begin{align}
    \left(x+\frac{1}{2x}\right)-\frac{1}{2}\ln\frac{x+1}{x-1}=2\frac{b_{\rm max}}{r_{\rm M}}.
\end{align}%
The integral (\ref{squareroot}) is
\begin{equation}
    \int\sqrt{1+4b^2/r_{\rm M}^2}\frac{db}{b}\sim 2\frac{b_{\rm max}}{r_{\rm M}}-\ln\frac{b_{\rm min}}{r_{\rm M}}-1.
\end{equation}
The integral over the impact parameters is then evaluated as
\begin{align}
    I_b&=\frac{1}{2}\int\frac{db}{b} +\frac{1}{2}\int\sqrt{1+4b^2/r_{\rm M}^2}\frac{db}{b} +\frac{1}{r_{\rm M}^2}\int b db= \cr
    &\sim\frac{1}{2}\ln\frac{b_{\rm max}}{b_{\rm min}}+\frac{b_{\rm max}}{r_{\rm M}}-\frac{1}{2}\ln\frac{b_{\rm min}}{r_{\rm M}}-\frac{1}{2}+\frac{b_{\rm max}^2}{2r_{\rm M}^2}=\cr
    &=\ln\Lambda+\frac{b_{\rm max}^2}{2r_{\rm M}^2}+\frac{b_{\rm max}}{r_{\rm M}}-\frac{1}{2}\ln\frac{b_{\rm max}}{r_{\rm M}}-\frac{1}{2}.
\end{align}
\section{Further details on the mean field approach}
\label{Kandrup calc}
For the system of $N+1$ particles considered in Section \ref{S3} described by the distribution function $\mathcal{F}(\mathbf{r}, \mathbf{p}; t)$ the average of a given phase-space observable $Q$, taken with respect to said distribution function is
\begin{equation}
    \langle Q\rangle_{\mathcal{F}}(t):=\int\mathcal{F}(\mathbf{r}, \mathbf{p}; t)Q(\mathbf{r}, \mathbf{p})d\Sigma,
\end{equation}%
where $d\Sigma=d^{3N}\mathbf{r}d^{3N}\mathbf{p}$ is the differential phase-space element.\\
\indent The initial configuration $\mathcal{F}_0$ is perturbed by the passage of the $(N+1)$-th particle, (our usual test particle). In virtue of the Third Law of Dynamics, this corresponds to the force decelerating the test particle once the force due to the unperturbed configuration $\mathcal{F}_0$ is subtracted:
\begin{equation}
\label{F fr}
    F_0^{fr}=\langle F_0^{tot}\rangle_{\mathcal{F}} -\langle F_0^{tot}\rangle_{\mathcal{F}_0}, \quad F_0^{tot}=-\sum_{i=1}^N F_i^{tot}.
\end{equation}%
The field particles are assumed statistically uncorrelated in their initial state $\mathcal{F}_0(\mathbf{r}, \mathbf{p})=\prod_{i=1}^N f(\mathbf{r}_i, \mathbf{p}_i)$. We also assume their velocity distribution to be a Maxwellian:
\begin{equation}
    f(\mathbf{r}_i, \mathbf{p}_i)=\mathcal{N}f_1(\mathbf{r}_i)e^{-\beta\mathbf{p}_i^2/2m_i},
\end{equation}%
with temperature $\beta$ and normalization $\mathcal{N}$. The distribution of the spatial coordinates is chosen such that the field particles are limited within a certain volume; therefore, it can be also expressed as $f_1(\mathbf{r}_i)=e^{-\beta m_i\Phi_i(\mathbf{r}_i)}$, with an efficient potential $\Phi_i$.\\
\indent For our purposes, we set $f_1$ as the limit of an infinite, homogeneous distribution of field particles. Such a homogeneous distribution on a finite volume $Vol\subseteq\mathbb{R}^3$ is given for $f_1(\mathbf{r_i})=\frac{1}{|Vol|}\chi_{Vol}(\mathbf{r}_i)$, namely, when $\Phi_i$ has the profile of a rigid box. The average number density of particles can be then defined as $n=N/|Vol|$. In the limit of an infinite homogeneous system, this parameters should be taken constant as $N, |Vol|\rightarrow\infty$. The unperturbed force $\langle F_0^{tot}\rangle_{\mathcal{F}_0}$ vanishes, so that the dynamical friction $F_0^{fr}$ can be obtained simply as $\langle F_0^{tot}\rangle_{\mathcal{F}}$.\\
\indent The field particles evolves following the time-dependent Hamiltonian:%
\begin{equation}
\label{Ham}
    H(\mathbf{r}, \mathbf{p}; t)=\sum_{i=1}^N \frac{\mathbf{p}_i^2}{2m_i} +\sum_{i=1}^N W_i(\mathbf{r}).
\end{equation}%
As discussed in Sect. (\ref{S3}), we cannot make the usual substitution of $W_i=\sum_{j\neq i}W_{ij}(|\mathbf{r}_i-\mathbf{r}_j|)$ with $W_{ij}(r)=-Gm_i m_j/r$. This is because,  in MOND, the superposition principle does not hold. The potentials should be rather expressed in using the QuMOND formalism (\ref{QuMOND}), so that:%
\begin{equation}
\label{Wi}
    -\nabla W_i=m_i\nu\left(\frac{|\mathbf{g}_{Ni}|}{a_0}\right)\mathbf{g}_{Ni}, \quad s.t. \quad -\sum_{j\neq i}\nabla W_{ij}=m_j\mathbf{g}_{Ni}.
\end{equation}
As a consequence, it is useless to define the forces between couples of particles $\mathbf{F}(j\rightarrow i)=-\partial_{\mathbf{r}_i}W_{ij}(|\mathbf{r}_i-\mathbf{r}_j|)=-\mathbf{F}(i\rightarrow j)$, given that the total QuMOND force on the $i-$th particle is not simply $\mathbf{F}_i^{tot}\neq\sum_{j\neq 0}\mathbf{F}(j\rightarrow i)$. The force $\mathbf{F}_i^{tot}$ can be calculated instead, only as
\begin{equation}
    \mathbf{F}_i^{tot}=-\partial_{\mathbf{r}_i}W_i(\mathbf{r}),
\end{equation}
where $W_i(\mathbf{r})$ is given by the non-linear QuMOND formula (\ref{Wi}). The evolution equation is therefore expressed as
\begin{equation}
\label{evol1}
    \dot{\mathcal{F}}=-\sum_{i=1}^N \frac{\mathbf{p}_i}{m_i}\cdot\frac{\partial\mathcal{F}}{\partial\mathbf{r}_i} -\sum_{i=1}^N \mathbf{F}_i^{tot}\cdot\frac{\partial\mathcal{F}}{\partial\mathbf{p}_i}:=-iL[\mathcal{F}].
\end{equation}%
The sums here arise from those in the Hamiltonian (\ref{Ham}). In practice, the meaning of Eq. (\ref{evol1}) is just that the position of each field particle evolves according to the velocity at each instant, as well as the momentum follows the total force.\\
\indent Following \cite{1983Ap&SS..97..435K}, it is useful to change the phase-space coordinates by taking the test particle's reference:
\begin{equation}
    \hat{\mathbf{r}}_i:=\mathbf{r}_i-\mathbf{R}(t)+\mathbf{R}(0), \quad \hat{\mathbf{p}}_i:=\mathbf{p}_i-\mathbf{P}(t).
\end{equation}%
The evolution operator for $\mathcal{F}$ hence takes the form%
\begin{equation}
\label{evol2}
    \dot{\mathcal{F}}(\hat{\mathbf{r}}_i, \hat{\mathbf{p}}_i; t)=-iL[\mathcal{F}]+\sum_{i=1}^N \dot{\mathbf{P}}(t)\cdot\frac{\partial\mathcal{F}}{\partial\mathbf{p}_i}:=-i\mathcal{L}[\mathcal{F}],
\end{equation}%
where $\dot{\mathbf{P}}$ acts as an apparent force on the $i$-th particle. Except for this, in this frame the forces do not change $\hat{F}_i^{tot}\equiv F_i^{tot}$, since even in MOND they depend only on the respective particles positions. Moreover, it is useful to define the perturbation $\tilde{\mathcal{F}}(t):=\mathcal{F}(t)-\mathcal{F}_0$ of the distribution function so that
\begin{equation}
    \langle Q\rangle_{\mathcal{F}}=\langle Q\rangle_{\mathcal{F}_0}+\langle Q\rangle_{\tilde{\mathcal{F}}}
\end{equation}%
for any quantity $Q$. This will be used for evaluating Eq. (\ref{F fr}), so that $\mathbf{F}_0^{fr}=\langle \mathbf{F}_0^{tot}\rangle_{\tilde{\mathcal{F}}}$. The distribution function $\tilde{\mathcal{F}}$ evolves as%
\begin{align} \label{evol3}
    \dot{\tilde{\mathcal{F}}}=-i\mathcal{L}[\tilde{\mathcal{F}}]+\beta\mathcal{F}_0(\mathbf{r}, \mathbf{p})\sum_{i=1}^N\left[\frac{\hat{\mathbf{p}}_i}{m_i}+\mathbf{V}(t)\right]\cdot\left[F_i^{tot}+m_i\frac{\partial\hat{\Phi}_i}{\partial\hat{\mathbf{r}}_i}(t)\right],
\end{align}%
where again $\hat{\mathbf{p}}_i/m_i+\mathbf{V}(t)=\mathbf{p}_i/m_i$, and the term $\partial\hat{\Phi}_i/\partial\hat{\mathbf{r}}_i$ vanishes in the infinite, homogeneous limit. In principle, keeping the contribution of $\hat{\Phi}$ would allow us to investigate in a self-consistent fashion the MOND EFE, see Sect. \ref{S3}. At this point, Eq. (\ref{evol3}) can be rewritten in its simpler form:
\begin{equation}
    \dot{\tilde{\mathcal{F}}}=-i\mathcal{L}[\tilde{\mathcal{F}}]+\beta\mathcal{F}_0\sum_{i=1}^N \mathbf{p}_i\cdot\mathbf{g}_i^{MOND}.
\end{equation}%
Its solution can thus be formally expressed as%
\begin{equation}
    \tilde{\mathcal{F}}(t)=\beta\mathcal{F}_0\int_0^t d\tau G_{\mathcal{L}}(\tau\rightarrow t)\left[\sum_{i=1}^N \mathbf{p}_i\cdot\mathbf{g}_i\right]
\end{equation}%
with the Greenian $G_{\mathcal{L}}$ of the operator $\mathcal{L}$.
After all the substitution we have a formal, general expression for the dynamical friction on the test particle in the form%
\begin{align} \label{C14}
    \mathbf{F}_0^{fr}&=\int d\Sigma \mathbf{F}_0^{tot}\tilde{\mathcal{F}}=\int d\Sigma \mathbf{F}_0^{tot}\beta\mathcal{F}_0\int_0^t d\tau G_{\mathcal{L}}(\tau\rightarrow t)\left[\sum_{i=1}^N \mathbf{v}_i\cdot\mathbf{F}^*_i\right]= \cr
    &=-\beta \int d\Sigma \mathcal{F}_0\left(\sum_{j=1}^N \mathbf{F}_i^{tot}\right) \int_0^t d\tau G_{\mathcal{L}}(\tau\rightarrow t)\left[\sum_{i=1}^N \mathbf{v}_i\cdot\mathbf{F}^*_i\right],
\end{align}%
where $\mathbf{F}^*_i=\mathbf{F}_i^{tot}+m_i\partial\hat{\Phi}_i/\partial\hat{\mathbf{r}}_i$ becomes $\mathbf{F}_i^{tot}$ in the infinite homogeneous system limit. Here we have exploited the fact that the full set of $N+1$ particles is a closed, classic dynamical system, so that $\sum_{i=0}^N \mathbf{F}_i^{tot}=0$.\\
\indent To perform the integrals appearing in Eqs. (\ref{C14}), we now make some simplifying hypotheses. First of all, we perform the infinite homogeneous limit. Second, if the system has only a finite memory, we can replace the integration of $\tau$ on $[0; t]$ with an integration of $[0; \infty)$. Moreover, we make use of the standard linear trajectory approximation, which allows us to simplify $G_{\mathcal{L}}(\tau\rightarrow t)[Q]\cong Q(t-\tau)$. Under these approximations, the DF force becomes
\begin{equation}
    \mathbf{F}_0^{fr}\cong -\beta \int d\Sigma \mathcal{F}_0\left(\sum_{j=1}^N \mathbf{F}_j^{tot}(t)\right) \int_0^{\infty} d\tau\left(\sum_{i=1}^N \mathbf{v}_i(t-\tau)\cdot\mathbf{F}_i^{tot}(t-\tau)\right).
\end{equation}%
This can be further simplified since $\langle F_i F_j\rangle=0$ for the off-diagonal combinations $i\neq j$, since the field particles are by definition statistically uncorrelated. We are then left with
\begin{align}
    \mathbf{F}_0^{fr}&\cong -\beta \int d\Sigma \mathcal{F}_0\sum_{i=1}^N \mathbf{F}_i^{tot}(t) \int_0^{\infty} d\tau \mathbf{v}_i(t-\tau)\cdot\mathbf{F}_i^{tot}(t-\tau)= \cr
    &=-\beta \int_0^{\infty} d\tau \left\langle\sum_{i=1}^N \mathbf{F}_i^{tot}(t)\left(\mathbf{v}_i(t-\tau)\cdot\mathbf{F}_i^{tot}(t-\tau)\right)\right\rangle_{\mathcal{F}_0}= \cr
    &=-\beta \int_0^{\infty} d\tau N\left\langle\mathbf{F}_1^{tot}(t)\left(\mathbf{v}_1(t-\tau)\cdot\mathbf{F}_1^{tot}(t-\tau)\right)\right\rangle_{\mathcal{F}_0}.
\end{align}%
We note that the last step is justified only if the $N$ field particles have all the same mass $m$, so that they are statistically equivalent in the initial phase-space distribution. Averaging $\mathcal{F}_0$ for the $i=1$ particle, practically means that only the $f(\mathbf{r}_1, \mathbf{p}_1)$ distribution function is relevant, and hence
\begin{align}   \left\langle\mathbf{F}_1^{tot}\left(\mathbf{v}_1\cdot\mathbf{F}_1^{tot}\right)\right\rangle_{\mathcal{F}_0}&=\int m\mathbf{g}_1^M\left(m\mathbf{v}_1\cdot\mathbf{g}_1^M\right) \prod_{i=1}^N f(\mathbf{r}_i, \mathbf{p}_i)d^{3N}\mathbf{r}d^{3N}\mathbf{p}= \cr
    &=m^2\int\mathbf{g}_1^M\left(\mathbf{v}_1\cdot\mathbf{g}_1^M\right) f(\mathbf{r}_1, \mathbf{p}_1)d^3\mathbf{r}_1 d^3\mathbf{p}_1= \cr
    &=m^2\int d^3\mathbf{r}_1 \int d^3\mathbf{v}_1 \frac{n}{N}f(\mathbf{v}_1) \mathbf{g}_1^M\left(\mathbf{v}_1\cdot\mathbf{g}_1^M\right)= \cr
    &=m^2\frac{n}{N}\left[\int f(\mathbf{v}) d^3\mathbf{v}\right]\left[\int (\mathbf{v}\cdot\mathbf{g}_1^M) \mathbf{g}_1^M d^3\mathbf{r}_1\right],
\end{align}%
where $f(\mathbf{v}_1)=\mathcal{N}e^{-\beta mv_1^2}$ is again the Maxwellian velocity distribution. The mean square speed is thus given by $\langle v^2\rangle=3/m\beta$ that can be substituted to obtain
\begin{align}
    \mathbf{F}_0^{fr}\cong&-\frac{3n}{m\langle v^2\rangle}m^2\int_0^{\infty} d\tau\left[\int f(\mathbf{v}) d^3\mathbf{v}\right]\left[\int (\mathbf{v}\cdot\mathbf{g}_1^M(t-\tau)) \mathbf{g}_1^M(t) d^3\mathbf{r}_1\right]=\cr
    \cong&-\frac{3\rho G^2 M^2}{\langle v^2\rangle}\int_0^{\infty}d\tau \int f(\mathbf{v}) d^3\mathbf{v} \cr
    &\times\int d^3\mathbf{s} \nu\left(\frac{r_{\rm M}^2}{|\mathbf{s}-\tilde{\mathbf{v}}\tau|^2}\right)\frac{\mathbf{v}\cdot(\mathbf{s}-\tilde{\mathbf{v}}\tau)}{|\mathbf{s}-\tilde{\mathbf{v}}_1\tau|^3} \nu\left(\frac{r_{\rm M}^2}{s^2}\right)\frac{\mathbf{s}}{s^3},
\end{align}%
since $\mathbf{g}_1^M(\mathbf{s}, t)=GM\nu\left(GM/a_0 s^2\right){\mathbf{s}}/{s^3}$ and $\mathbf{s}(t-\tau)\cong\mathbf{s}-\tilde{\mathbf{v}}_1\tau$, being $\tilde{\mathbf{v}}=\mathbf{v}-\mathbf{V}$.
\section{Numerical methods}\label{nummeth}
The dynamics of a particle confined by a static potential $\Phi$ under the combined effect of force fluctuations and friction is given by the Langevin-type equations
\begin{equation}\label{langevin}
\frac{{\rm d}^2\mathbf{r}}{{\rm d}t^2}=-\nabla\Phi_{\mathbf{r}}-\eta_{\mathbf{r},\mathbf{v}}\mathbf{v}+\delta F;\quad\mathbf{v}=\frac{{\rm d}\mathbf{r}}{{\rm d}t}.
\end{equation}
In the most general case, the DF coefficient $\eta$ depends explicitly on the position and velocity through the phase-space distribution of the embedding system (also sourcing the potential $\Phi$) and $\delta F$ is a fluctuating force per unit mass with distribution and amplitude connected to $\eta$ via a fluctuation-dissipation relation. For the systems of interest (Newtonian or MOND) we solve Eq. (\ref{langevin}) using the \cite{2004PhRvE..69d1107M} scheme
\begin{eqnarray}\label{mannella}
x^\prime=x(t+\Delta t/2)=x(t)+\frac{\Delta t}{2}v(t)\nonumber\\
v(t+\Delta t)=c_2\left[c_1v(t)+\Delta t \nabla\Phi(x^\prime)+d_1 \tilde F(x^\prime) \right]\nonumber\\
x(t+\Delta t)=x^\prime+\frac{\Delta t}{2}v(t+\Delta t),
\end{eqnarray}
written here for simplicity for a 1D system and fixed time step $\Delta t$, where the coefficients $c_1,c_2$ and $d_1$ are given by 
\begin{equation}
c_1=1-\frac{\eta\Delta t}{2};\quad c_2=\frac{1}{1+\eta\Delta t/2};\quad d_1=\sqrt{2\zeta\eta\Delta t}.
\end{equation}
In the equations above $\tilde F$ is sampled from a norm 1 Gaussian and $\zeta$ in the case of a delta correlated noise is fixed by the standard deviation of the distribution of $F$ as
\begin{equation}\label{sd}
\langle F(x,t) F(x,t^\prime)\rangle=2\eta\zeta\delta(t-t^\prime).
\end{equation}
In practice, if such distribution is unknown, we would then, following \cite{2000MNRAS.311..719K}, assume $\zeta=\mathcal{E}$ where $\mathcal{E}$ is the instantaneous relative (positive) energy per unit mass of the test particle along the orbit. In the simulations discussed in this paper the contribution of the fluctuating term is artificially set to 0 as it would be negligible for the type of gravitational systems under consideration. We note that if $\nu=\zeta=0$ Eqs. (\ref{mannella}) simply become the standard Verlet second order scheme in the drift-kick-drift form.
\end{appendix}
\end{document}